\documentclass[11pt,a4paper,english,twoside]{article}

\usepackage{a4wide}
\usepackage{amssymb, amsmath}
\usepackage{graphicx}
\usepackage[all]{xy}
\usepackage{hyperref}
\hypersetup{linktocpage}
\usepackage{enumerate}
\usepackage{xcolor}
\usepackage{dsfont}
\usepackage{empheq}
\usepackage{cite}
\usepackage{float}

\usepackage{soul}
\usepackage{subcaption}

\newcommand{\beq}{\begin{equation}}
\newcommand{\eeq}{\end{equation}}
\def\bea#1\eea{\begin{align}#1\end{align}}
\def\beal#1\eeal{\begin{subequations}\begin{align}#1\end{align}\end{subequations}}
\newcommand{\nn}{\nonumber}

\def\del {\partial}
\def\d {{\rm d}}

\begin{document}
\numberwithin{equation}{section}

\begin{titlepage}

\begin{flushright}
CERN-TH-2018-251
\end{flushright}

\begin{center}

\phantom{DRAFT}

\vspace{2.8cm}

{\LARGE \bf{Further refining the de Sitter swampland conjecture}}\\

\vspace{2 cm} {\Large David Andriot$^{1}$ and Christoph Roupec$^{2}$}\\
 \vspace{0.9 cm} {\small\slshape $^1$ Theoretical Physics Department, CERN\\
1211 Geneva 23, Switzerland}\\
 \vspace{0.2 cm} {\small\slshape $^2$ Institute for Theoretical Physics, TU Wien\\
Wiedner Hauptstrasse 8-10/136, A-1040 Vienna, Austria}\\
\vspace{0.5cm} {\upshape\ttfamily david.andriot@cern.ch; christoph.roupec@tuwien.ac.at}\\

\vspace{3cm}

{\bf Abstract}
\vspace{0.1cm}
\end{center}

\begin{quotation}
\noindent
We propose an alternative refined de Sitter conjecture. It is given by a natural condition on a combination of the first and second derivatives of the scalar potential. We derive our conjecture in the same weak coupling, semi-classical regime where the previous refined de Sitter conjecture was derived, using the same tools together with a few more assumptions that we discuss. We further test and constrain free parameters in our conjecture using data points of a classical type IIA supergravity setup. Interestingly, our conjecture easily accommodates slow-roll single field inflation with a concave potential, favored by observations. The standard quintessence potential is in tension with our new conjecture, and we thus propose a different type of quintessence model.

\end{quotation}

\end{titlepage}

\newpage

\tableofcontents

\section{Introduction: de Sitter swampland conjectures}

The swampland program is an interesting recent development for the phenomenology of quantum gravity theories, as it concretely connects to bottom-up approaches, and to questions regarding effective field theories and their U.V.~completions. The swampland is the space of effective theories that cannot be coupled consistently to a theory of quantum gravity; more concretely, this includes the set of phenomenological models which cannot be derived as a low energy effective theory of a quantum gravity. The swampland program then amounts to propose a list of explicit criteria that a theory should (or not) obey if it sits in the swampland: see \cite{Brennan:2017rbf} for a review. This provides concrete and powerful tools to discriminate among various phenomenological models, such as cosmological or BSM models, which would otherwise appear on equal footing. The difficulty in the program is to establish these criteria on solid grounds: while some of them are based on pure quantum gravity arguments involving e.g.~black holes, or tested to a non-trivial extent in some string theory context, others are more debated. This is the case for the de Sitter conjecture \cite{Obied:2018sgi} that led to important discussions. Inspired by the difficulty of getting well-controlled de Sitter vacua from string theory (see \cite{Danielsson:2018ztv} for a review), the criterion \eqref{conj1} detailed below has been proposed \cite{Obied:2018sgi}: one of its implications is to forbid any de Sitter solution in a theory outside the swampland. Important concerns have then been raised regarding known examples of de Sitter local maxima, among e.g.~classical perturbative de Sitter string backgrounds (see \cite{Andriot:2018ept} for a review, and \cite{Roupec:2018mbn} for some criticism) or particle physics scalar potentials (Higgs, QCD axion, neutral pion) \cite{Denef:2018etk, Murayama:2018lie, Choi:2018rze, Hamaguchi:2018vtv}.\footnote{Another interesting argument was given in \cite{Conlon:2018eyr}, regarding the existence of a de Sitter maximum in the effective theory obtained after compactification on a Calabi-Yau manifold with a single K\"ahler modulus. Nothing however ensures that this maximum appears consistently within the low energy effective theory, meaning below a cutoff scale such as e.g.~the Kaluza--Klein scale. In particular, since the argument uses the asymptotic behavior of the potential, this maximum could appear only at large values of the K\"ahler modulus, where the Kaluza--Klein scale would indeed be low.} This situation has led to a call for a refined de Sitter conjecture \cite{Andriot:2018wzk, Garg:2018reu}, which would only forbid de Sitter minima but allow maxima, i.e.~unstable (tachyonic) de Sitter solutions: such a criterion has been formulated recently \cite{Ooguri:2018wrx} (see also \cite{Garg:2018reu}) in the form of \eqref{conj0} detailed below. This refined de Sitter conjecture has been tested with various phenomenological models \cite{Wang:2018kly, Fukuda:2018haz, Lin:2018rnx, Agrawal:2018rcg, Chiang:2018lqx, Cheong:2018udx}, and has been discussed in relation to stringy constructions \cite{Olguin-Tejo:2018pfq, Garg:2018zdg, Blaback:2018hdo, Heckman:2018mxl, Blanco-Pillado:2018xyn, Junghans:2018gdb, Emelin:2018igk, Banlaki:2018ayh} or in a more general swampland context \cite{Hebecker:2018vxz, Dvali:2018jhn, Schimmrigk:2018gch, Ibe:2018ffn}. In the present paper, we propose an alternative refined de Sitter conjecture.\\

We consider a four-dimensional (4d) theory of real scalar fields $\phi^i$ coupled to gravity, whose dynamics is governed by a scalar potential $V(\phi^j)$, with an action given by
\beq
S=\int_4 \d^4 x \sqrt{|g_4|} \left(\frac{{M_p}^2}{2}\, {\cal R}_4 - \frac{1}{2} g_{ij} \del_{\mu} \phi^i \del^{\mu} \phi^j - V \right) \ , \label{action}
\eeq
where $g_{ij}(\phi^k)$ is the field space metric, $M_p$ is the 4d Planck mass, and the 4d space-time index $\mu$ is raised and lowered with the 4d metric of signature $(-,+,+,+)$. Such a model could serve various phenomenological purposes, such as describing multi-field cosmological inflation. We introduce the following quantities and notations
\bea
& |\nabla V| = \sqrt{g^{ij} \del_{i} V \del_{j} V} \ ,\ {\rm min}\ \nabla \del V =\mbox{minimum eigenvalue of}\  g^{ik} \nabla_k \del_j V \ ,\nn\\
\mbox{and for}\ V > 0 \ ,\qquad & \epsilon_V = \frac{{M_p}^2}{2} \left(\frac{|\nabla V|}{V}\right)^2 \ ,\ \eta_V= {M_p}^2\, \frac{{\rm min}\ \nabla \del V}{V} \ ,
\eea
where $\del_{i}$ stands for $\del_{\phi^i}$ and $\nabla_i$ for the corresponding covariant derivative in field space. The matrix $M^i{}_j= g^{ik} \nabla_k \del_j V$ is the mass (square) matrix for constant fields or constant fluctuations; see e.g.~\cite{Achucarro:2010da, Renaux-Petel:2015mga} for the non-constant case. The eigenvalues of this matrix then give the mass spectrum, and whether their minimum is negative tells us if a tachyon is present. The parameters $\epsilon_V$ and $\eta_V$ match the slow-roll parameters of single-field inflation, and are also used for multi-field inflation, even though different ones, $\epsilon_H$ and $\eta_H$, may then be more relevant (see e.g.~\cite{Hetz:2016ics}). Our conventions match those of \cite{Hertzberg:2007wc, Flauger:2008ad}, of relevance to the present work.

The refined de Sitter conjecture of \cite{Ooguri:2018wrx}, already present for most of it in \cite{Garg:2018reu}, applies to models of the form \eqref{action}. It is stated as follows:\footnote{To be precise, the condition \eqref{conj2} is rather written in \cite{Ooguri:2018wrx} considering the Hessian in an orthonormal frame, instead of the present mass matrix. Here we consider this more covariant formulation. We recall that the set of eigenvalues of a matrix is not changed when applying $GL$ transformations on it; in other words, ${\rm min}\ \nabla \del V$ is unchanged when transforming our covariant $M^i{}_j$ by field space diffeomorphisms.} {\it an effective theory for quantum gravity, i.e.~not in the swampland, should verify either}
\begin{subequations}\label{conj0}
\bea
|\nabla V| \geq \frac{c}{M_p} V \ ,&\qquad \qquad {\rm with}\ c \simeq 1  \label{conj1}\\
{\rm {\it or}}\qquad {\rm min}\ \nabla \del V \leq - \frac{c'}{{M_p}^2} V \ ,&\qquad \qquad {\rm with}\ c' \simeq 1 \label{conj2}
\eea
\end{subequations}
The inequalities can be rewritten for $V>0$ as
\beq
\sqrt{2 \epsilon_V} \geq c \quad {\rm {\it or}} \quad \eta_V \leq -c' \ .
\eeq
The first condition \eqref{conj1} corresponds to the original de Sitter conjecture proposed in \cite{Obied:2018sgi}. It implies the absence of any de Sitter solution in the model \eqref{action}, i.e.~any positive extremum of the potential $V|_0 >0$, giving rise to a 4d de Sitter space-time with ${\cal R}_4 = 4/{M_p}^2\, V|_0$. Including the second derivative of the potential to distinguish between de Sitter maxima and minima was proposed in a first refinement in \cite{Andriot:2018wzk}, and another formulation was given in \cite{Garg:2018reu, Ooguri:2018wrx} in the form of the second condition \eqref{conj2}. Thanks to the logical ``or'', de Sitter maxima are now allowed, and the counterexamples to \eqref{conj1} initially put forward are now accommodated \cite{Ooguri:2018wrx}.

The formulation of this conjecture \eqref{conj0} is somewhat peculiar: it is given by two distinct conditions \eqref{conj1} and \eqref{conj2} on two different quantities $\epsilon_V$ and $\eta_V$. On top of easily singling out de Sitter maxima, one reason for this formulation is the derivation of this conjecture proposed in \cite{Ooguri:2018wrx} (more details on it are provided in Section \ref{sec:deriv}). There, the condition \eqref{conj1} is derived in a weak coupling, semi-classical regime, where the latter is defined in particular by the requirement $\eta_V \gtrsim -1$, hence the separation between the two quantities and conditions. As a consequence, this formulation does not provide information on both quantities simultaneously. As such, this is different than the refined conjecture proposed in \cite{Andriot:2018wzk}: the latter was formulated as a single inequality on a linear combination of the potential, its first and second derivatives, that would again only allow de Sitter maxima and forbid minima. A criticism against that proposal was that it was not formulated covariantly with respect to field space diffeomorphisms,\footnote{As explained in \cite{Andriot:2018wzk}, the idea behind this formulation was to have a sufficient condition (hence not necessarily covariant) to conclude on a tachyon, using Sylvester's criterion.} except in the single field case. This problem is straightforward to solve by writing the same inequality in terms of the previous covariant quantities: this corresponds to the inequality \eqref{conj} below with $q=1$. The resulting condition has however a more important caveat, not much noticed so far: it is not trivial in the limit $M_p \rightarrow \infty$. This limit is actually a crucial aspect of the swampland program, as it corresponds to the decoupling of field theory from gravity. The swampland conjectures constrain effective theories that couple consistently to quantum gravity, so they should become empty, meaning trivial, when gravity decouples. It is the case of condition \eqref{conj1}, which thus assures, thanks to the logical ``or'', that the whole conjecture \eqref{conj0} is trivially satisfied in this limit.\\

Building on this discussion, there is a natural de Sitter conjecture to be written, that is expressed as a single condition on both $\epsilon_V$ and $\eta_V$. We propose the following statement as an alternative refined de Sitter conjecture: {\it a low energy effective theory of a quantum gravity that takes the form \eqref{action} should verify, at any point in field space where $V>0$,}
\beq
\left(M_p\, \frac{|\nabla V|}{V}\right)^q - a\, {M_p}^2\, \frac{{\rm min}\nabla \del V}{V} \geq b \ ,\qquad \qquad {\rm with}\ a+b=1 \ ,\ a,b>0\ , \ q > 2 \label{conj}
\eeq
This inequality can also be written, in terms of the slow-roll parameters, as
\beq
(2 \epsilon_V)^{\frac{q}{2}} - a\, \eta_V \geq b \ .
\eeq
We conjecture in this paper that \eqref{conj} holds in general. This is an interesting and concrete claim to test, and we provide some evidence for it. As discussed in Section \ref{sec:deriv}, it is nevertheless not excluded that the set of values of the three real parameters $q,a,b$ is broader than proposed here, and we could keep this in mind for future investigations.

\begin{figure}[H]
\begin{center}
\begin{subfigure}[H]{0.4\textwidth}
\includegraphics[width=\textwidth]{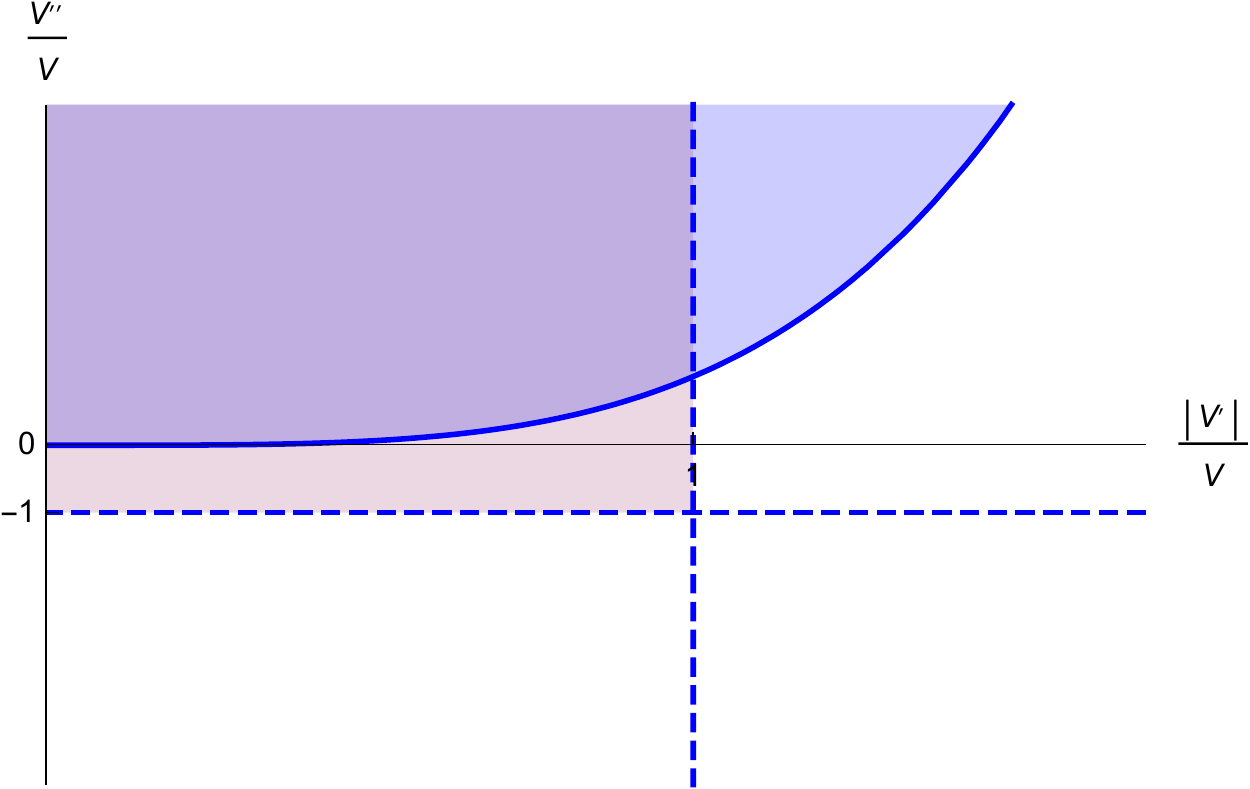}\caption{$a\simeq1$}\label{fig:abig}
\end{subfigure}
\qquad
\begin{subfigure}[H]{0.4\textwidth}
\includegraphics[width=\textwidth]{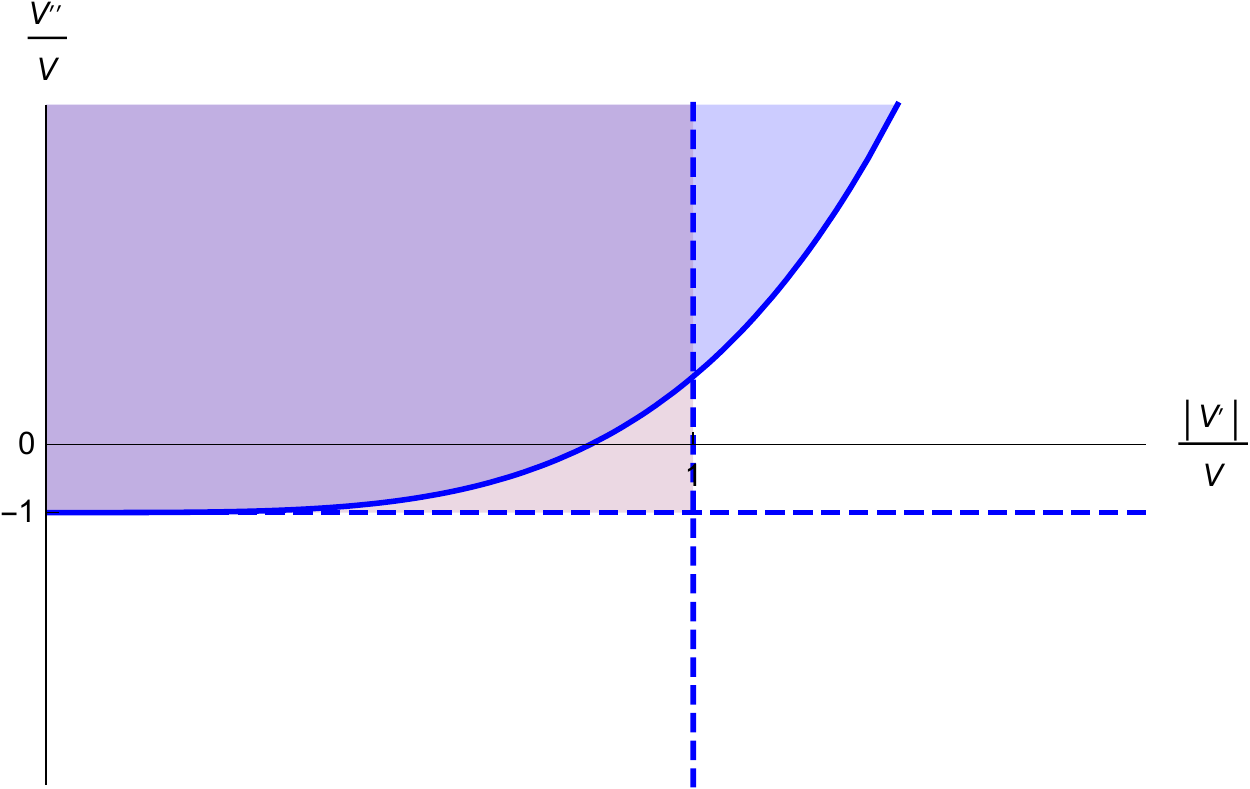}\caption{$a=\tfrac{1}{2}$}\label{fig:aav}
\end{subfigure}
\begin{subfigure}[H]{0.4\textwidth}
\includegraphics[width=\textwidth]{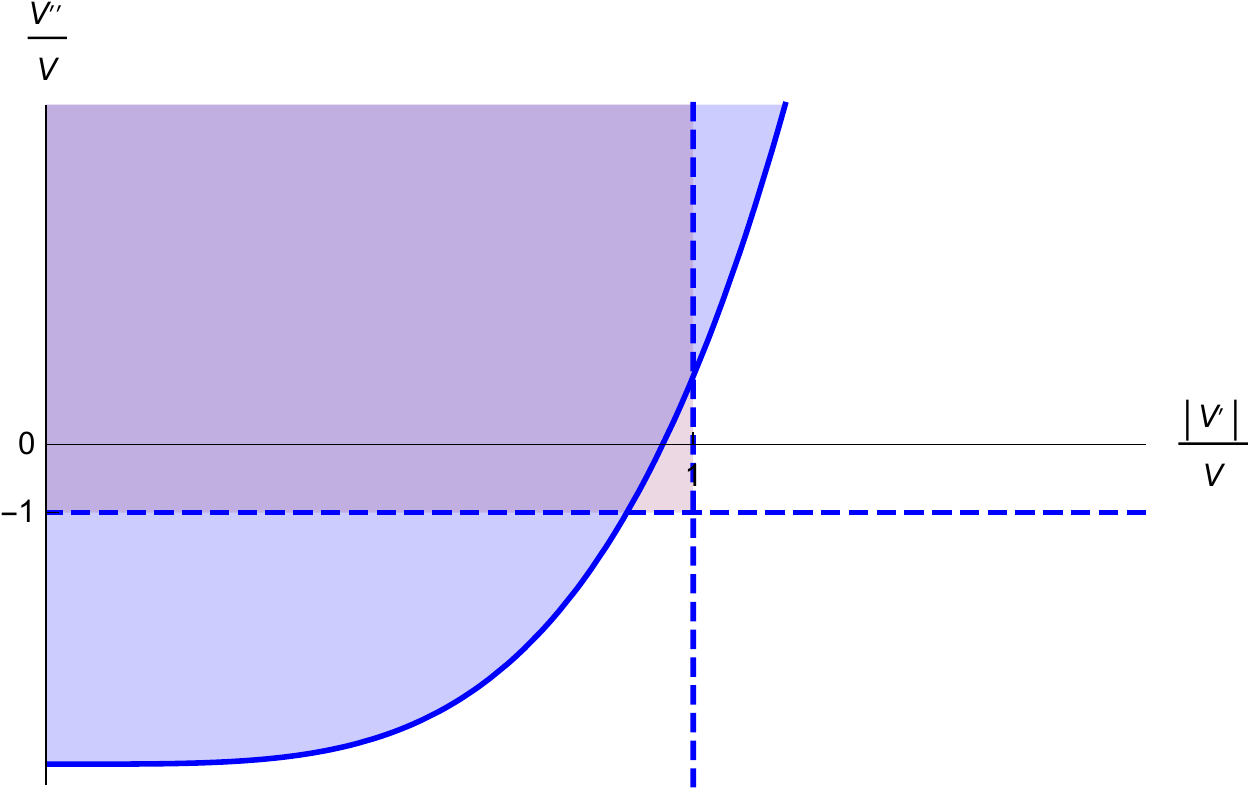}\caption{$a=\tfrac{1}{5.7}$}\label{fig:asmall}
\end{subfigure}
\caption{Parameter space $(\sqrt{2\epsilon_V},\eta_V)$ denoted symbolically $\left(\tfrac{|V'|}{V},\tfrac{V''}{V}\right)$, as probed by the two different refined de Sitter conjectures: the swampland regions for each of them is colored. The pink region, delimited by the dashed lines, is excluded by the conjecture \eqref{conj0}, while the blue region, delimited by the plain line, is excluded by the new conjecture \eqref{conj}. For this illustration, we take $c=c'=1$, $q=4$ and let $a$ vary.}\label{fig:3excl}
\end{center}
\end{figure}

In the limit $M_p \rightarrow \infty$, the condition \eqref{conj} becomes $(|\nabla V|/V)^q  \geq 0$, which is trivially satisfied, as required. To reach this, one should first divide the condition by $(M_p)^q$ so that the Planck mass appears in the denominator, as for the other conjecture \eqref{conj0}, and have the strict inequality $q>2$. Note also that for both conjectures, this limit makes sense when considered over all field space, and not at particular points. In addition to having an appropriate decoupling limit, one has that the positivity condition $a,b>0$ leads to $({\rm min}\nabla \del V|_0)/(V|_0) < 0$ at an extremum of the potential, which only allows for tachyonic de Sitter solutions and forbids de Sitter vacua. These are also two aspects of the refined conjecture \eqref{conj0}, so the difference between the two is more about the values of parameters and bounds. We summarize this in Figure \ref{fig:3excl}. While both conjectures allow for large values of $(\epsilon_V , |\eta_V|)$ and forbid small values of $\epsilon_V$ with positive $\eta_V$, they differ in two to three regions of this parameter space. One conjecture is therefore not stronger than the other, i.e.~one is not implying the other, they simply agree on most of parameter space, and do not in these regions where alternatively one is stronger than the other. A consequence is nevertheless that they define different boundaries to the swampland, with potentially important implications for cosmological models, as studied in Section \ref{sec:cosmo}; in particular, slow-roll single-field inflation is allowed by the new conjecture \eqref{conj} in the case of Figure \ref{fig:abig}, while it cannot be accommodated by the previous one \eqref{conj0}.

We provide in Section \ref{sec:deriv} a derivation of the new conjecture \eqref{conj} in the same weak coupling, semi-classical regime as the one introduced in \cite{Ooguri:2018wrx}. Our derivation uses the same tools as those for \eqref{conj0}, together with few more assumptions that we discuss. This derivation provides the interesting relation $a+b=1$. In this regime, both the condition \eqref{conj1} and the new conjecture \eqref{conj} should thus hold, which places us in the top right quarter of Figures \ref{fig:abig} - \ref{fig:asmall}. There, the new conjecture \eqref{conj} is actually stronger, and can thus be viewed as a further refinement to the previous \eqref{conj0}, in the form of additional information involving both $\epsilon_V$ and $\eta_V$.

In Section \ref{sec:check}, we test the new conjecture \eqref{conj} against some data points $(\epsilon_V, \eta_V)$, obtained in the 10d classical and perturbative string framework of type IIA supergravity with $O_6$ orientifold planes and orbifolds \cite{Danielsson:2011au}. These points are all obtained at large volume and small string coupling; we discuss explicitly other possible caveats regarding consistency with string theory \cite{Roupec:2018mbn}. This data provides interesting bounds on the parameters $q,a$. We study in Section \ref{sec:cosmo} cosmological implications of this conjecture. Its preference for concave potentials, in remarkable agreement with observational constraints on inflation \cite{Akrami:2018odb}, is problematic for the usual quintessence potential. We then propose different quintessence models in agreement with our conjecture \eqref{conj}. We summarize our results in Section \ref{sec:ccl} and make further comments. In the remainder of the paper, we set $M_p=1$ for simplicity.

\section{Derivation in the weak coupling regime}\label{sec:deriv}

In \cite{Ooguri:2018wrx}, the condition \eqref{conj1} is derived in a weak coupling regime, corresponding to going at any large distance in the space of low energy scalar fields. Going to such asymptotic distances is argued to correspond to parametrically controlled regimes of string theory, as is for instance the case of a small string coupling with a large negative dilaton or small $\alpha'$ corrections with a large volume. This large field space distances also allow to use the swampland distance conjecture \cite{Ooguri:2006in}, generalized and refined to the case of a non-zero scalar potential \cite{Baume:2016psm, Klaewer:2016kiy}. That conjecture indicates an increase in the number of light degrees of freedom below a certain cutoff, when going to this asymptotic regime. This counting of microscopic degrees of freedom is then related to the notion of entropy. The relevant one here is the Gibbons-Hawking entropy \cite{Gibbons:1977mu} in a de Sitter space-time. The derivation of \cite{Ooguri:2018wrx} also uses the Bousso bound \cite{Bousso:1999xy} on the entropy, that is argued to be saturated in this weak coupling regime. Those concepts are applied there more generally to an accelerating universe. To be allowed to use these concepts, one should remain within a semi-classical description: to that end, it is argued that $\eta_V$ should not be smaller than $-1$, hence the second condition \eqref{conj2}. In this section, we derive our conjecture \eqref{conj} using precisely the same tools, therefore making this analysis in the same weak coupling, semi-classical, regime. More precisely, we show that parameters $q,a,b$ can easily be found compatible with the requirements of the conjecture \eqref{conj}, even though we cannot strictly exclude here some other values. For this reason, the derivation made in this section should only be viewed as an argument in favor of the conjecture \eqref{conj}, in this regime.\\

Let us consider the potential $V$ of a single, canonically normalized, scalar field $\phi$ of a low energy effective theory of a quantum gravity. Using the distance conjecture, the Gibbons-Hawking entropy and the (saturated) Bousso bound, it is argued in \cite{Ooguri:2018wrx} as summarized above that this potential should be given as follows, in the regime discussed previously
\beq
V(\phi) = \left( n(\phi)\, e^{d \phi} \right)^{-k} \ ,\quad d,k > 0 \ , \label{pot}
\eeq
where the function $n$ counts the number of towers of states becoming light, given a certain cutoff scale.\footnote{Our constant $d$ was denoted $b$ in \cite{Ooguri:2018wrx}, and $k$ was given there by $\tfrac{2\gamma}{2-\delta}$.} We thus have $n > 0$ and $\del_{\phi} n = n'\geq 0$ in this regime \cite{Ooguri:2018wrx}, i.e.~the number of contributing towers grows when going asymptotically. From \eqref{pot}, one derives
\beq
\frac{V'}{V}=-k\left( d + \frac{n'}{n}\right) \ ,\label{vpv}
\eeq
from which one deduces with $n'\geq 0$ that
\beq
\frac{|V'|}{V} \geq k d = c \ . \label{VpVineq}
\eeq
This provides the condition \eqref{conj1} proposed in \cite{Obied:2018sgi, Ooguri:2018wrx}, with the constant $c$.

To reach the conjecture \eqref{conj}, we first compute
\beq
\frac{V''}{V} = k^2 d^2 + \frac{n'}{n} k \left( \frac{k+1}{n} + 2kd  \right) - k \frac{n''}{n} \ .
\eeq
We then assume the following
\beq
n \geq 1 \ ,\ \frac{n''}{n} \geq 0 \ .
\eeq
Since $n$ is growing, it starts in this regime with a minimum. This function counts the number of towers contributing, so it looks reasonable to take this minimum bigger or equal to $1$. In addition, following the idea of the distance conjecture, it appears natural that more and more towers will contribute when going to the asymptotics, hence $n''\geq 0$ and the above assumption.\footnote{One may prefer relaxing the assumption $n''\geq 0$ by rather introducing a negative minimum to $n''$: indeed, $n''$ should not be too negative, otherwise the sign of $n'$ could quickly change, in contradiction with $n'\geq 0$. If $n''$ admits a finite minimum $n_0''$ and $n$ is growing, the ratio $\tfrac{n_0''}{n}$ will at some point in the asymptotics become negligible compared to the constant $k d^2$ in $\tfrac{V''}{V}$. Then, we can work in this later regime; in that case, the rest of the reasoning leads eventually to the same result as when assuming $n''\geq 0$.} These two assumptions may however be discussed further, indicating the possibility of small corrections to the final values of the parameters $a,b$ that we derive. With these extra assumptions with respect to \cite{Ooguri:2018wrx}, one deduces
\beq
\frac{V''}{V} \leq k^2 d^2 + \frac{n'}{n} k \left( k+1 + 2kd  \right)  \ . \label{VppVineq}
\eeq
Coming back to \eqref{vpv}, one has
\beq
\left(\frac{|V'|}{V}\right)^q = k^q \left( d + \frac{n'}{n}\right)^q \geq k^q \left( d^q + q \frac{n'}{n} d^{q-1} \right) \ , \label{expansion}
\eeq
where the inequality is obtained because all terms in the power expansion are positive.\footnote{In the case where $n'$ were bounded and $n$ large enough, one could also view this expansion as an approximation in powers of $\frac{n'}{n}$.} While this expansion is obvious for an integer $q$, it holds more generally for a real value (at least for $q>2$) thanks to the following relations for real positive $x,y$
\beq
(x+y)^q - x^q = \int_x^{x+y} \d t\ q\, t^{q-1}  \geq \int_x^{x+y}  \d t\ q\, x^{q-1} = q\, y\, x^{q-1}\ .
\eeq
We now combine \eqref{VppVineq} and \eqref{expansion} to get
\beq
\left(\frac{|V'|}{V}\right)^q - a\, \frac{V''}{V}  \geq k^q  d^q - a  k^2 d^2 + \frac{n'}{n} \left( q k^q d^{q-1} - a k \left( k+1 + 2kd  \right) \right) \ .
\eeq
Eliminating the (a priori non-constant) piece in $\frac{n'}{n} $, we fix $a$, and deduce
\beq
\left(\frac{|V'|}{V}\right)^q - a\, \frac{V''}{V}  \geq b \ ,\quad \quad {\rm with}\ a= \frac{q\, (kd)^{q-1}}{\left( k+1 + 2kd  \right)} \ ,\ b=  (k d)^q\,  \frac{k+1 + kd (2-q)}{\left( k+1 + 2kd  \right)} \ . \label{result}
\eeq
The reasoning of \cite{Ooguri:2018wrx} leading to \eqref{VpVineq} was meant to justify the condition \eqref{conj1}, i.e.~one should have
\beq
c=kd \simeq 1 \ .
\eeq
We choose this value as well; this may also be viewed as fixing a freedom in parameterizing exponents in the asymptotic behavior of the potential \eqref{pot}. This gives
\beq
a= \frac{q}{k+3} \ ,\quad b= 1 - \frac{q}{ k + 3 } \ ,\quad a+b=1 \ .\label{result2}
\eeq
This reproduces the conjecture \eqref{conj} as long as we impose $q < k+3$, which is certainly possible for $q>2$, e.g.~with $q=3$ or more generally some $\mathbb{R}$-valued $q$.\footnote{Note that different, more constrained, values can be obtained for $a,b$, still satisfying $a+b=1$, if one chooses inequalities larger than \eqref{expansion}. Those would however have less chances to be saturated.} This concludes our derivation of \eqref{conj}.

\section{Testing the new conjecture}\label{sec:check}

We would like to test the new conjecture \eqref{conj} within a string theory framework. To that end, we need a setup where the derivation of the potential $V$ is established on solid grounds, while points with $V>0$ can be reached. Obtaining de Sitter solutions from string theory in a well-controlled manner is however a hard task, a situation which has motivated the de Sitter swampland conjectures. Schematically, when restricting to classical and perturbative string backgrounds, de Sitter solutions are rare, while going beyond this framework provides more ingredients and enhances the chances to find such solutions, but it implies at the same time a reduced control on the construction. More precisely, non-perturbative or quantum corrections have often been used to produce de Sitter vacua, but their contributions are typically added at a 4d effective level. Having such semi-classical or quantum contributions under control in a 10d string theory framework is by definition difficult, and this often leads to debate about these constructions (see e.g.~\cite{Moritz:2017xto, Sethi:2017phn, Gautason:2018gln} and \cite{Cicoli:2018kdo, Kachru:2018aqn, Akrami:2018ylq}). Classical perturbative 10d solutions do not suffer from the same doubts, but they are still plagued with issues, at a different level. The problems there are due to the way any such 10d solution is found: one typically solves 10d supergravity equations, as these theories are low energy effective theories for string theories in such classical regimes. Once a solution is found, one should verify that it satisfies certain requirements that places it in this appropriate regime; if not, it remains a solution of supergravity, without being embeddable in string theory, or at least not as a classical perturbative background. This is, as for the other constructions, a swampland type of problem, even though the consistency requirements are not on the same level. The most important requirements for supergravity solutions to ensure their stringy consistency would be to have a large internal volume ``${\rm vol}$'' (compared to the string scale) and a small string coupling $g_s=e^{\phi}$ fixed by the dilaton $\phi$. Further requirements include flux quantization that we disregard in the following, and a bounded number of orientifolds that we come back to. When moving away from an extremum of the potential, i.e.~adding dynamics, while still having $V>0$ (these are also points of interest to test our conjecture), there is no reason to believe that these difficulties get eased. All these consistency requirements makes it hard to find a single string theory setup with $V>0$ that can be fully trusted. In other words, it is difficult to find an appropriate string setup to test our conjecture.

The advantage of the 10d classical (supergravity) framework over other approaches is that consistency issues with string theory are easily identified. We thus propose here to test the de Sitter refined conjectures with $(\epsilon_V, \eta_V)$ data points coming from such a framework, while indicating explicitly possible caveats. Doing such a test has at least the advantage of illustrating what could be possible constraints on these conjectures and their parameters.

\begin{itemize}
  \item Classical de Sitter solutions
\end{itemize}

Despite being rare, a few classical de Sitter solutions have been found in 10d type IIA supergravity on 6d group manifolds, with $O_6$ orientifolds and various orbifolds \cite{Caviezel:2008tf, Flauger:2008ad, Danielsson:2009ff, Caviezel:2009tu, Danielsson:2010bc, Danielsson:2011au, Roupec:2018mbn}; all of them are unstable. Stringy consistency issues for all these solutions have however been pointed-out in \cite{Roupec:2018mbn, Junghans:2018gdb, Banlaki:2018ayh}; this appears in agreement with the idea that no de Sitter solution could be found in a parametrically controlled regime of string theory \cite{Ooguri:2018wrx}. In particular, a difficulty was to get at the same time a large ${\rm vol}$, a small $g_s$, and a small number of $O_6$. The latter corresponds to the number of fixed points of the geometry and is therefore typically of order $1$, at least in these settings; this has to be contrasted with the number of $D_6$-branes which does not admit such a bound. We still use here the same framework, based on \cite{Danielsson:2011au}.\footnote{While our conventions on $\epsilon_V, \eta_V$ match those of \cite{Hertzberg:2007wc, Caviezel:2008tf, Flauger:2008ad, Danielsson:2011au}, we note that the dilaton $e^{- 4 D}$ of \cite{Flauger:2008ad} corresponds to the following quantities of the other references: $D=-\hat{\tau}/\sqrt{2}$ and $e^{- 2D}=\tau^2 = {\rm vol}\, e^{- 2 \phi}$.} Our various data points, coming from \cite{Roupec:2018mbn}, correspond to different group manifolds and underlying algebras, as well as different orbifolds, which in turn select different flux components. For each of these possibilities, our numerical approach minimizes $\epsilon_V$, sometimes reaching a numerical $0$, thus obtaining a de Sitter solution. We present our twenty most relevant points in Table \ref{tab:ourpoints}. In terms of consistency requirements, we encounter the difficulties mentioned above. One typically has the freedom of using various rescalings to bring some of the relevant quantities to appropriate values, but not all of them. Here we obtain a large ${\rm vol}$ and small $g_s$, while leaving the number of $O_6$ unbounded; we could have equivalently restricted to at most one $O_6$ (for each of the four sets of sources directions, or equivalently four tadpoles),\footnote{In this setting, one has $D_6$ and $O_6$ along four different sets of directions. As shown in \cite{Andriot:2017jhf}, a sum of these sources contributions has to be negative to get a de Sitter solution, requiring as usual the presence of some $O_6$ \cite{Maldacena:2000mw}, but this does not need to hold for each of the four sets separately.} but then we would not have obtained a small $g_s$. We refer to section 4 of \cite{Roupec:2018mbn} for more details.

\begin{itemize}
  \item Classical non-extrema with $V>0$
\end{itemize}

Interestingly, non-vanishing values of $(\epsilon_V, \eta_V)$ were looked for in \cite{Blaback:2013fca} in a similar setup. There, both $\epsilon_V$ and $\eta_V$ were minimized. The four best points were given explicitly and we copy them in Table \ref{tab:Danpoints}. Precisely the same type of consistency issues were encountered in that work: the scaling freedom was used to obtain large ${\rm vol}$ and small $g_s$, while leaving the number of $O_6$ unbounded.

Our data points also provided us with non-extrema: our best points are given by $(\sqrt{2\epsilon_V}, \eta_V)$ being equal to $(1.633, 0.6650)$ and $(1.776, -0.3330)$. Those points are however not relevant to constrain the conjectures, so we do not include them in the following. It is interesting to note that the boundary curve for the new conjecture \eqref{conj} always passes through the point $(1,1)$, for any allowed value of the parameters. Therefore, any point $\sqrt{2\epsilon_V} \geq 1, \ \eta_V \leq 1$ does not provide any constraint.\\

\begin{table}
\begin{center}
\begin{tabular}{|c|c|}\hline

$\sqrt{2 \epsilon_V}$ & $\eta_V$ \\\hline

$3.738\cdot 10^{-10}$& 	$-2.495$ \\
$3.711\cdot10^{-7}$&	$-3.663$ \\
$1.479\cdot10^{-6}$&	$-3.673$ \\
$7.733\cdot10^{-6}$&	$-3.727$ \\
$4.487\cdot10^{-6}$&	$-3.904$ \\
$2.712\cdot10^{-6}$&	$-3.914$ \\
$8.251\cdot10^{-7}$&	$-3.915$ \\
$1.105\cdot10^{-6}$&	$-3.915$ \\
$9.719\cdot10^{-7}$& 	$-3.918$ \\
$1.248\cdot10^{-6}$&	$-3.922$ \\
$1.728\cdot10^{-6}$&	$-3.927$ \\
$1.894\cdot10^{-6}$&	$-3.964$ \\
$2.378\cdot10^{-6}$&	$-3.973$ \\
$1.129\cdot10^{-6}$&	$-4.049$ \\
$2.859\cdot10^{-6}$&	$-4.049$ \\
$6.255\cdot10^{-6}$&	$-4.145$ \\
$1.998\cdot10^{-6}$&	$-4.186$ \\
$4.634\cdot10^{-6}$&	$-4.187$ \\
$4.289 \cdot 10^ {-5}$&	$-4.211$ \\
$2.562 \cdot 10^ {-5}$&	$-4.297$ \\\hline

\end{tabular} \caption{20 data points of classical de Sitter solutions} \label{tab:ourpoints}
\end{center}
\end{table}

\begin{table}
\begin{center}
\begin{tabular}{|c|c|}\hline

$\sqrt{2 \epsilon_V}$ & $\eta_V$ \\\hline
$ 0.950$  & $-0.077$ \\
$ 0.927$  & $-0.151$ \\
$ 0.875$  & $-0.162$ \\
$ 0.885$  & $-0.318$ \\\hline

\end{tabular} \caption{4 data points of classical non-extrema from \cite{Blaback:2013fca}} \label{tab:Danpoints}
\end{center}
\end{table}

\begin{figure}[H]
\begin{center}
\includegraphics[width=12.0cm]{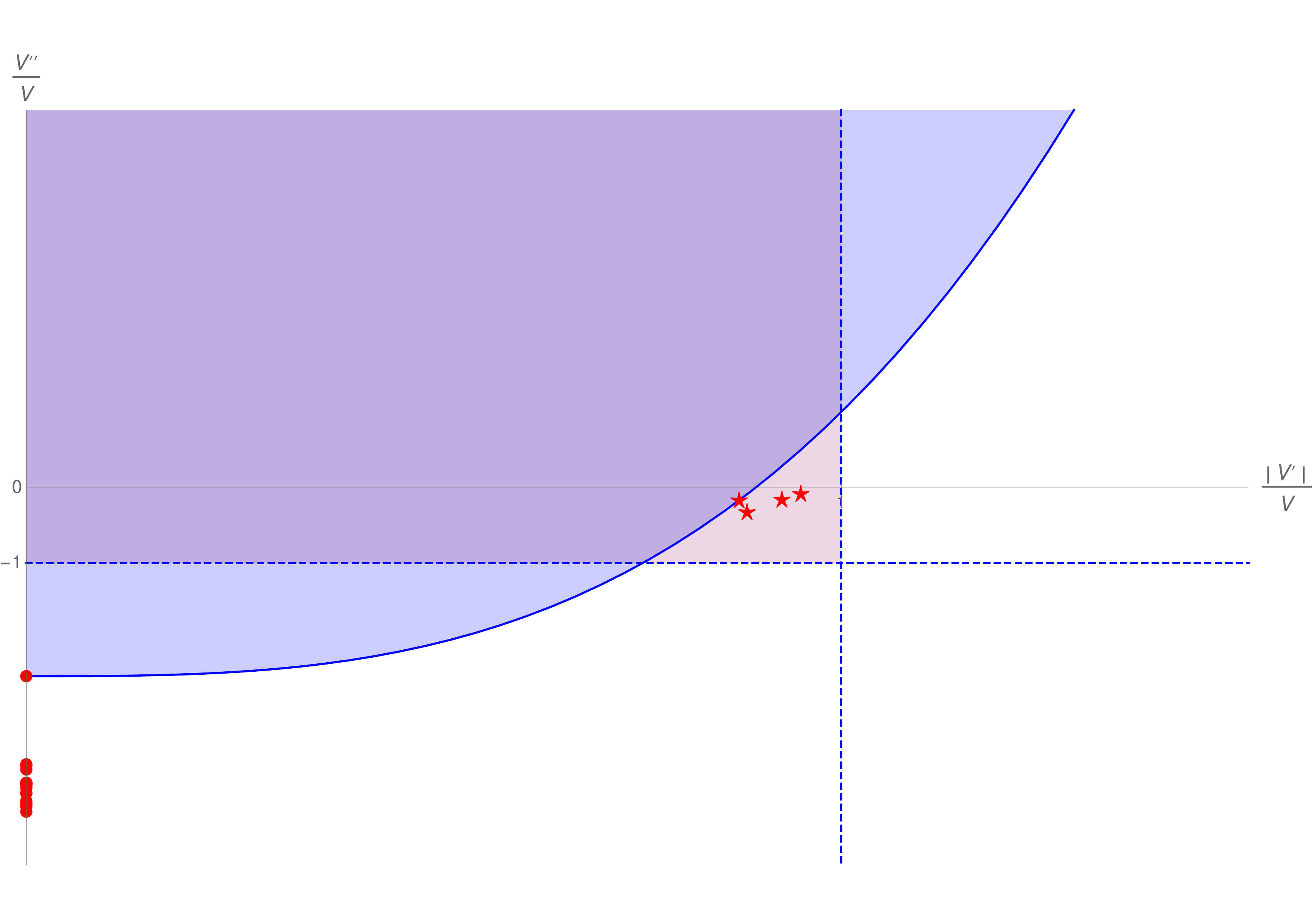}\caption{Parameter space $(\sqrt{2 \epsilon_V},\eta_V)$ with swampland regions delimited by the two different de Sitter conjectures \eqref{conj0} and \eqref{conj}, together with data points of classical type IIA supergravity models. Points of Table \ref{tab:ourpoints} are dots on the left axis, while points of Table \ref{tab:Danpoints} are stars in the central region. The boundary curve for the new conjecture \eqref{conj} is here bounded by these data points, providing constraints on the parameters $q,a$.}
\label{Fig:data}
\end{center}
\end{figure}

We now use data points of Table \ref{tab:ourpoints} and \ref{tab:Danpoints} to test the conjectures: they provide interesting constraints, as illustrated by Figure \ref{Fig:data}. None of these points strongly disagrees with the conjecture \eqref{conj0}: the points of Table \ref{tab:Danpoints} from \cite{Blaback:2013fca} can be considered consistent with \eqref{conj1} if one allows for the usual flexibility of $c$ being of order $1$, instead of strictly $1$. But these data points provide constraints on the parameters $q,a$ of the new conjecture \eqref{conj}. From the points $\epsilon_V=0$ with $\eta_V<0$, one deduces the constraint $a\geq \frac{1}{1-\eta_V}$: here we obtain $a \geq 0.286$. If we take this boundary value for $a$, the curve gets bounded by the point $(0.875, -0.162)$ of Table \ref{tab:Danpoints}, as in Figure \ref{Fig:data}, which provides an upper bound $q\leq 3.03$. The latter is the most conservative value: choosing a higher value for $a$ allows to raise the upper bound on $q$. Given that we require $q>2$, the conservative upper bound $3.03$ is still very interesting.

We close this section by mentioning few other data points found in the literature. In the same type IIA setup of \cite{Caviezel:2008tf, Flauger:2008ad} are mentioned de Sitter solutions with $\eta_V \lesssim -2.4$; more precisely, points with $\eta_V \simeq -3.7$ and $\eta_V \simeq -2.5$ are given in \cite{Flauger:2008ad}. In type IIB supergravity, the only classical de Sitter solutions obtained \cite{Caviezel:2009tu} gave $\eta_V \simeq -3.1$. In \cite{Garg:2018zdg} are obtained points that reach the bound $\eta_V \lesssim -10$, with a possibility of an extra factor of order $1$. These values do not appear more constraining than what we presented here; rather the fact they are of the same order is interesting. Finally, an interesting upper bound of $\eta_V \leq -4/3$ was analytically derived in 4d ${\cal N}=1$ supergravity for specific de Sitter critical points, parametrically close to a no-scale Minkowski solution \cite{Junghans:2016abx} (see also \cite{Covi:2008ea, Covi:2008cn, Junghans:2016uvg}). While this value would be a little more constraining than the data points we used, it is for now not clear to us whether this upper bound can be reached, especially through a consistent string construction.

\section{Cosmological consequences: concave or convex potentials}\label{sec:cosmo}

We study in this section the consequences of the new conjecture \eqref{conj} for the early and late universe cosmological models; we consider for simplicity canonically normalized scalar fields. The conjecture \eqref{conj} is much easier to satisfy with $\eta_V \leq 0$. For single-field models with $V>0$, this means that concave potentials are favored by this conjecture over convex ones. Regarding the early universe and single-field inflation models, this is interesting because remarkably, the latest results by the Planck Collaboration \cite{Akrami:2018odb} (see Fig. 6, 7 or 8 there) now indicate a strong preference for concave potentials. Regarding the late universe, the de Sitter swampland conjectures typically propose to replace the cosmological constant by a quintessence model, as an answer to the absence of de Sitter vacuum \cite{Agrawal:2018own}. A standard potential for a single-field quintessence model is $V(\phi)=V_0\, e^{-\lambda \phi}$ with $\lambda >0$, which is a convex potential. Current observational bounds set $\lambda \lesssim 0.6$ \cite{Agrawal:2018own} (see \cite{Heisenberg:2018yae} for slightly different bounds), and we will see that this is not in agreement with our conjecture. However, there is a priori nothing forbidding a concave quintessence potential. In the future though, a purely concave potential would lead to a big crunch scenario; it could then as well turn back later to a convex potential, and we illustrate below this proposal with a concrete potential. One may also wonder about multi-field quintessence models; all these scenarios could accommodate our conjecture. We now make a more precise analysis.

\begin{itemize}
  \item Single-field inflation
\end{itemize}

It is simple to verify that the values $a \ll 1$, $b\simeq 1$ or $a=b=\tfrac{1}{2}$ do not allow for parameters $\epsilon_V$ and $\eta_V$ in a slow-roll regime. However, the values $a\simeq 1$, $b \ll 1$, as in Figure \ref{fig:abig}, can accommodate slow-roll inflation. For instance, a concave potential with $\epsilon_V \leq - \eta_V \simeq b \ll 1 $, as favored by current observational bounds \cite{Akrami:2018odb}, satisfies our conjecture \eqref{conj} $\forall q$. Moreover, for $q\simeq 2$ which remains a limiting case, the conjecture would even be satisfied for a convex potential with $\epsilon_V \simeq \eta_V \simeq b \ll 1 $. Matching observationally preferred (slow-roll single-field) inflation models is therefore made possible by the conjecture \eqref{conj}, provided $a\simeq 1$, $b \ll 1$. This is an important difference with the previous refined de Sitter conjecture \eqref{conj0}, which requires different, more involved models, such as \cite{Achucarro:2018vey, Kehagias:2018uem}.

\begin{itemize}
  \item Single-field quintessence
\end{itemize}

We first consider the potential $V(\phi)=V_0\, e^{-\lambda \phi}$ with $\lambda >0$. The conjecture \eqref{conj} then requires $\lambda^q -a \lambda^2 - b \geq 0$. The analysis of this condition is simple: remarkably, the inequality is saturated at $\lambda=1$, $\forall a,b$, i.e.~with a potential in $e^{-\phi}$. One further verifies that the inequality is satisfied only for $\lambda \geq 1$, crucially using that $q/(2a) > 1$. As mentioned previously, this is therefore not in agreement with the observational bounds.\footnote{While the conjecture \eqref{conj0} allows for some flexibility when considering $\lambda=0.6$ to be of order $1$, and therefore considers the condition \eqref{conj1} to be satisfied, the same flexibility is not enough in \eqref{conj} because of the powers of $\lambda$.} One option would then be to start with a concave potential, in agreement with current observational bounds, corresponding to our late universe so far, and branch it later on a decreasing exponential with a steep enough slope, if one wants to avoid a big crunch. On more general grounds, one may still want to have an asymptotic potential going as a $e^{-\lambda \phi}$ with $\lambda \geq 1$, either because of the Dine-Seiberg argument \cite{Dine:1985he}, the potential upper bound obtained in \cite{Hebecker:2018vxz}, or the asymptotic exponential decay \cite{Ooguri:2018wrx} expressed in \eqref{pot} and the condition \eqref{VpVineq}. We now make such a proposal concrete.

\begin{figure}[H]
\begin{center}
\includegraphics[width=10.5cm]{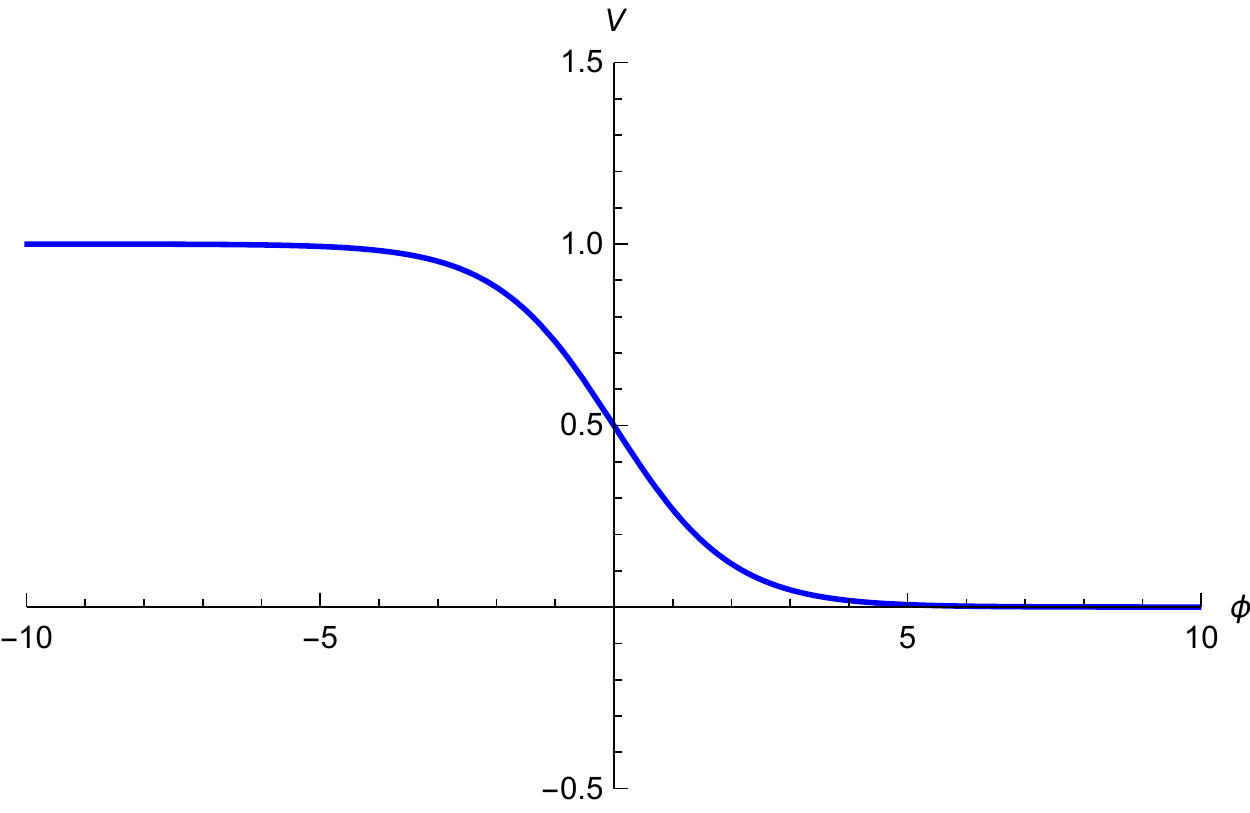}\caption{Potential \eqref{potquint} with $V_0=1,\ \lambda=\tfrac{1}{2}$.}\label{Fig:pot}
\end{center}
\end{figure}

Consider the following potential,\footnote{Interestingly, a similar potential appeared in \cite{Akrami:2017cir} in the context of $\alpha$-attractors, where the plateau was there rather used for inflation. We also note that similar shapes were discussed in \cite{Parameswaran:2016qqq}.} depicted in Figure \ref{Fig:pot}
\beq
V(\phi)=\frac{V_0}{2}(1-\tanh(\lambda \phi))= \frac{ V_0\, e^{-\lambda \phi}}{e^{\lambda \phi} + e^{-\lambda \phi}} \ , \qquad V_0, \lambda > 0\ .\label{potquint}
\eeq
\noindent On the left side, between $\phi_0$ and a point closer to the origin, say e.g.~$\phi_0=-10$ until $\phi=-3$, one can have (slow-roll) quintessence, and the potential is concave. One computes in general
\beq
\frac{|V'|}{V}= \frac{2\lambda e^{\lambda \phi}}{e^{\lambda \phi} + e^{-\lambda \phi}} \ ,\quad \frac{V''}{V}= 4\lambda^2 e^{\lambda \phi} \frac{e^{\lambda \phi} - e^{-\lambda \phi}}{(e^{\lambda \phi} + e^{-\lambda \phi})^2} \ ,
\eeq
and in the limit $\phi\rightarrow - \infty$, one gets $|V'|/V \sim 2 \lambda e^{2\lambda \phi}  $, $V''/V \sim -4 \lambda^2 e^{2\lambda \phi} $. We deduce that $\epsilon_V$ and $\eta_V$ are exponentially small, hence the slow-roll regime. In this regime, one has $w+1 = \frac{2}{3} \epsilon_V$ \cite{Agrawal:2018own}, which provides $w \sim -1$ to a much better precision than with the standard exponential potential that gives $\epsilon_V= \lambda^2/2$. This quintessence model is therefore in agreement with observations.

\begin{figure}[H]
\begin{center}
\includegraphics[width=10.5cm]{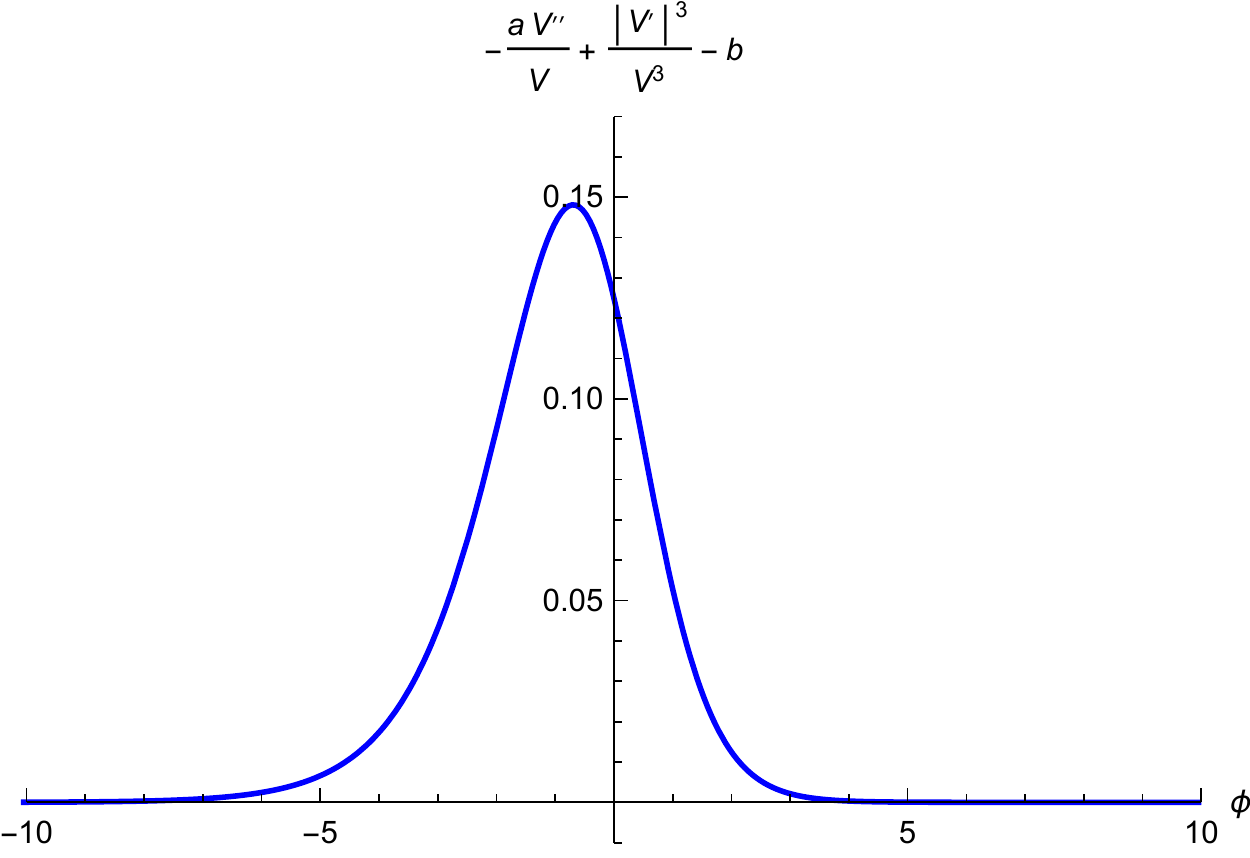}\caption{Check of the conjecture \eqref{conj} for the potential \eqref{potquint} where $\lambda = \tfrac{1}{2}$, with parameters \eqref{par} and $q=3$.}\label{Fig:potineq}
\end{center}
\end{figure}

We now show that we can find parameters $a,b,q$ such that the conjecture \eqref{conj} is satisfied for this potential, hence showing that a quintessence model can be built in agreement with our conjecture. First, as mentioned earlier, the inequality is saturated for a potential of the form $e^{-\phi}$. When $\phi\rightarrow +\infty$, $V \sim e^{-2\lambda \phi}$, so we fix for simplicity $\lambda = \tfrac{1}{2}$. On the $-\infty$ side, $\eta_V$ contribution is dominant over $\epsilon_V$, due to the $q$ power. Thanks to concavity, it is then easy to satisfy the inequality: it is enough to fix
\beq
b=\frac{1}{2}\times 4 \lambda^2 e^{2\lambda \phi_0} = \frac{1}{2} e^{-10} \ ,\ a=1-b \simeq 1 \ ,\label{par}
\eeq
where we keep in mind that quintessence has started at a finite time in the past, here related to $\phi_0$. The inequality remains satisfied from $-\infty$ until the origin since the potential is concave there. It however becomes convex afterwards and there could be a risk for the inequality if the $\epsilon_V$ contribution becomes to small: since $\epsilon_V < 1$, it means we should not have $q$ too big. We verify that $q=3$ is fine, as depicted in Figure \ref{Fig:potineq}.

The quintessence potential \eqref{potquint} is of course highly tuned. It should be viewed as an illustration that concave quintessence can be accommodated, and would be in agreement with our conjecture \eqref{conj}. Interestingly, we note that both this quintessence model and the observationally favored inflation models satisfy the new conjecture for parameters $a\simeq 1$, $b \ll 1$ as in Figure \ref{fig:abig}, the former requiring in addition $q$ not much bigger than $3$.

\section{Summary and outlook}\label{sec:ccl}

We propose in this paper a natural de Sitter swampland conjecture \eqref{conj}, which forbids de Sitter vacua and becomes trivial when decoupling from gravity. In the weak coupling, semi-classical regime where the previous refined de Sitter conjecture \eqref{conj0} was derived \cite{Ooguri:2018wrx}, we derive our new conjecture, given a few more assumptions that we discuss in Section \ref{sec:deriv}. In this regime, corresponding to the top right corner of parameter space in Figure \ref{fig:3excl}, the new conjecture \eqref{conj} appears stronger than the previous one \eqref{conj0} and can thus be viewed as a further refinement. This is not so surprising: we essentially obtain more information by computing $V''$ in this regime and comparing it to $V'$. The new conjecture \eqref{conj} depends on free real parameters $a>0,\ b=1-a>0,\ q>2$, on which we derive various constraints. In Section \ref{sec:check}, we use various data points $(\epsilon_V, \eta_V)$, obtained in a classical type IIA supergravity setup, to get the bound $a\geq 0.286$, corresponding to an upper bound $q \leq 3.03$; a higher value for $a$ raises the upper bound on $q$. This data should however be used with care, and we discuss explicitly its possible caveats in terms of consistency with string theory (essentially a matter of number of $O_6$ orientifolds). We study in Section \ref{sec:cosmo} the cosmological implications of the new conjecture, and find interestingly that it can easily accommodate slow-roll single field inflation with the parameters $a\simeq 1, b\ll 1$. The conjecture also favors concave potentials which is in good agreement with observational constraints \cite{Akrami:2018odb} on single field inflation. This last point is however more challenging for the usually considered quintessence potential: we discuss this and propose a different type of quintessence potential, in agreement with \eqref{conj}. Interestingly, that potential prefers again the parameters $a\simeq 1, b\ll 1$, also in agreement with the constraints obtained from the supergravity data points. The case of Figure \ref{fig:abig} is therefore favored. We also note that $2 < q \lesssim 3$ is a range that appears with the data points, the quintessence potential, and in the derivation where one eventually asks for $q<k+3$; we however keep in mind that each time, this range is not a strict bound. It would be interesting to test the new conjecture \eqref{conj} further, using in particular the particle physics models, such as the Higgs, the QCD axion, or the pion potentials.

A few new constructions of classical de Sitter solutions could, if their stringy consistency was improved, become challenging to any refined de Sitter conjecture. In \cite{Blaback:2018hdo}, a de Sitter solution was obtained with $\eta=0$, in a 4d supergravity. It does not involve any non-geometric flux, but only the standard RR-fluxes, the NSNS $H$-flux and geometric fluxes. In that sense, it has chances to admit a 10d description as a classical compactification. The novelty however is that the Jacobi identity on geometric fluxes is not satisfied. This is the same as the purely geometric Bianchi identity, which can be reformulated with the Riemann tensor as ${\cal R}^a{}_{[bcd]}=0$; its violation can be justified by the presence of a Kaluza--Klein monopole (see e.g.~\cite{Andriot:2014uda}, especially equation (3.5) there). This is however not systematic: sources (and fluxes) have to be quantized, sources of opposite charge could also be required due to compactness, and a 10d realisation would require warp factors. The simultaneous presence of $O_6$ orientifolds makes it difficult to hope for an explicit 10d realisation. It remains true that Kaluza--Klein monopoles appear to be a promising ingredient to get de Sitter solutions \cite{Silverstein:2007ac}. Another ingredient thought to be helpful are anti-branes, and an interesting classical de Sitter vacuum with $\overline{D}_6$ was obtained in \cite{Kallosh:2018nrk}. The latter however had small volume and large string coupling. More work is thus required to turn these recent interesting examples into trustworthy string constructions.

Having allowed de Sitter maxima by refining the de Sitter conjecture, but still forbidding minima, one may now wonder whether counterexamples of positive vacua can be found. The first examples that come to mind are again particle physics models, but also the de Sitter minimum often required for post inflation reheating. Thermal effects may however play a role in bringing these vacua to effectively positive value. It would be interesting to study whether these examples or others would challenge the refined de Sitter conjectures.

\vspace{0.4in}

\subsection*{Acknowledgements}

We warmly thank D.~Klaewer, M.~Kleban, C.~Ruef, S.~Sibiryakov and T.~Wrase for very useful exchanges on this project. D.~A.~thanks the Technical University of Vienna, the Centre de Physique des Houches, and the University of Liverpool, as well as all the members of these groups, for their warm hospitality during the completion of this project. The work of C.~R.~is supported by an FWF grant with the number P 30265.

\newpage

\providecommand{\href}[2]{#2}\begingroup\raggedright\endgroup

\end{document}